\documentclass[aps,preprint,superscriptaddress]{revtex4}
\usepackage{graphicx}
\usepackage{amsmath}
\usepackage{ulem}
\usepackage{bm}
\usepackage{CJK}
\usepackage{pdfpages}
\usepackage{color}
\usepackage{hyperref}
\begin{document}
\begin{CJK}{GBK}{song}
\title{Rethinking the solar flare paradigm}
\author{D. B. MELROSE\footnote{Email address: donald.melrose@sydney.edu.au}}
\affiliation{\rm SIfA, School of Physics, The University of Sydney, NSW 2006, Australia}

\date{\today}

\begin{abstract}
It is widely accepted that solar flares involve release of magnetic energy stored in the solar corona above an active region, but existing models do not include the explicitly time-dependent electrodynamics needed to describe such energy release. A flare paradigm is discussed that includes the electromotive force (EMF) as the driver of the flare, and the flare-associated current that links different regions where magnetic reconnection, electron acceleration, the acceleration of mass motions and current closure occur. The EMF becomes localized across regions where energy conversion occurs, and is involved in energy propagation between these regions.

\end{abstract}

\maketitle

\vskip 5mm
\textbf{1  Introduction}
\vskip 5mm

It is widely accepted that solar flares are due to explosive release (over tens of minutes) of magnetic energy that has been built up and stored above an active region in the corona over a longer time (days to weeks). The released energy appears in various forms of kinetic energy: mass motions, energetic particles and heat. Such release of magnetic energy is referred to as a magnetic explosion \cite{Moore2001,Melrose2012}. However, most models for solar flares do not include an essential feature of the physics of a magnetic explosion: the driver of a flare involves explicitly time-dependent electrodynamics.  The history of models for solar flares, over the decades since Giovanelli's 1948 model \cite{Giovanelli1948}, is well summarized by an archive of the cartoons used to describe each model (\url{http://solarmuri.ssl.berkeley.edu/~hhudson/cartoons/}). As discussed in Section~2, most of these models may be included in one of three classes: CSHKP models, circuit models or quadrupolar models, none of which is explicitly time-dependent. The missing ingredients are the electromotive force (EMF) and the flare-associated current. The time-changing coronal magnetic field implies a large-scale inductive electric field, whose line-integral along any closed path is interpreted as an EMF, $\Phi$, which tends to drive a flare-associated current along a closed path that needs to be identified in a specific flare model. The EMF is either ignored or replaced by a proxy, such as a photospheric dynamo, obscuring the essential role it plays as the driver of the flare-associated current. The EMF is the electrodynamic driver of the flare in the sense that it drives the current that links widely-separated regions, with the power released is due to the rate work is done by the EMF against the flare-associated current in localized regions along this current path. Note that although both an inductive electric field and a parallel electric field (associated with finite conductivity) appear in MHD models for flares \cite{ShibataM2011}, these are local fields, whereas the EMF and the current that it drives are on a global scale, and the parallel electric field arises when the EMF is localized along the path of the driven current. MHD, which is needed to model local regions along the flare-driven current path, needs to be complemented by the global  electrodynamics associated with the EMF and the flare-associated current that it drives. It is this global perspective that is obscured in many models for solar flares.

In discussing the energetics of solar flares, it is important to distinguish between the energy going into mass motions and that going into energetic electrons in the impulsive phase. A subset of flares is ``eruptive'' in the sense that the flare is associated with a coronal mass ejection (CME) that carries away a large fraction of the energy released. All flares have an ``impulsive'' phase in which a large fraction of the energy released goes into 10--20$\,$keV electrons: the electrons that precipitate into the denser regions of the solar atmosphere produce hard X-ray bursts (HXBs) \cite{Brown1971,Brown1976} and H$\alpha$ brightenings, and the electrons that escape produce type~III solar radio bursts. Explicit time-dependence is included in some magnetohydrodymanic (MHD) models for the acceleration of a CME \cite{Gibson1998,Wu2001,Chen2011}, but this is not the same as including the EMF as the driver of the flare. In the discussion of the energetics in this paper, emphasis is given to the acceleration of energetic electrons, as a defining feature of the impulsive phase of any flare. The general problem of the acceleration of solar energetic particles has been studied extensively \cite{Fletcher2011,Zharkova2011}, with most acceleration mechanisms applying to a small fraction of already suprathermal particles. In early literature, the acceleration of the electrons that produce HXBs was called ``bulk energization'' to indicate it involves all the electrons in a localized region having their mean energy increased by a large factor \cite{Ramaty1980,Benz1987,Melrose1990}. It has since been confirmed that the number density of the precipitating electrons in HXBs is comparable with the ambient electron number density \cite{Krucker2011}. More recent observations \cite{Janvier2014} suggest that the precipitating electrons are confined to very narrow current channels where they are accelerated by a parallel electric field \cite{Haerendel2017a,Haerendel2017b}. A flare model needs to include this bulk energization as the main form of dissipation for the released magnetic energy.

The neglect of the EMF and the flare-associated current obscures some important questions that need to be answered in a realistic time-dependent flare model: What is the flare-associated current and along what path does it flow? Where does the flare-associated current close across field lines? Where does the EMF become localized along the current path? Where does magnetic reconnection occur and what role does it play in the global electrodynamics? Where is the dissipation/acceleration region located? How is the magnetic energy transported along the current path to the acceleration/dissipation region? 

Existing flare models are classified and discussed briefly in Section~2. The requirements of a time-dependent flare model are discussed in Section~3, and some remarks on the acceleration/dissipation are given in Section~4. The conclusions are summarized in Section~5.

\vskip 5mm
\textbf{2  Existing flare models}	
\vskip 5mm

Most flare models can be classified as ``standard'' (or CSHKP)  models, circuit models or quadrupolar models. How the explicit time-dependence is avoided in each of these classes is discussed in this Section. 

\vskip 5mm
\textbf{2.1  CSHKP models}
\vskip 5mm

\begin{figure}[htbp]
\centering
\includegraphics[width = 8.5cm]{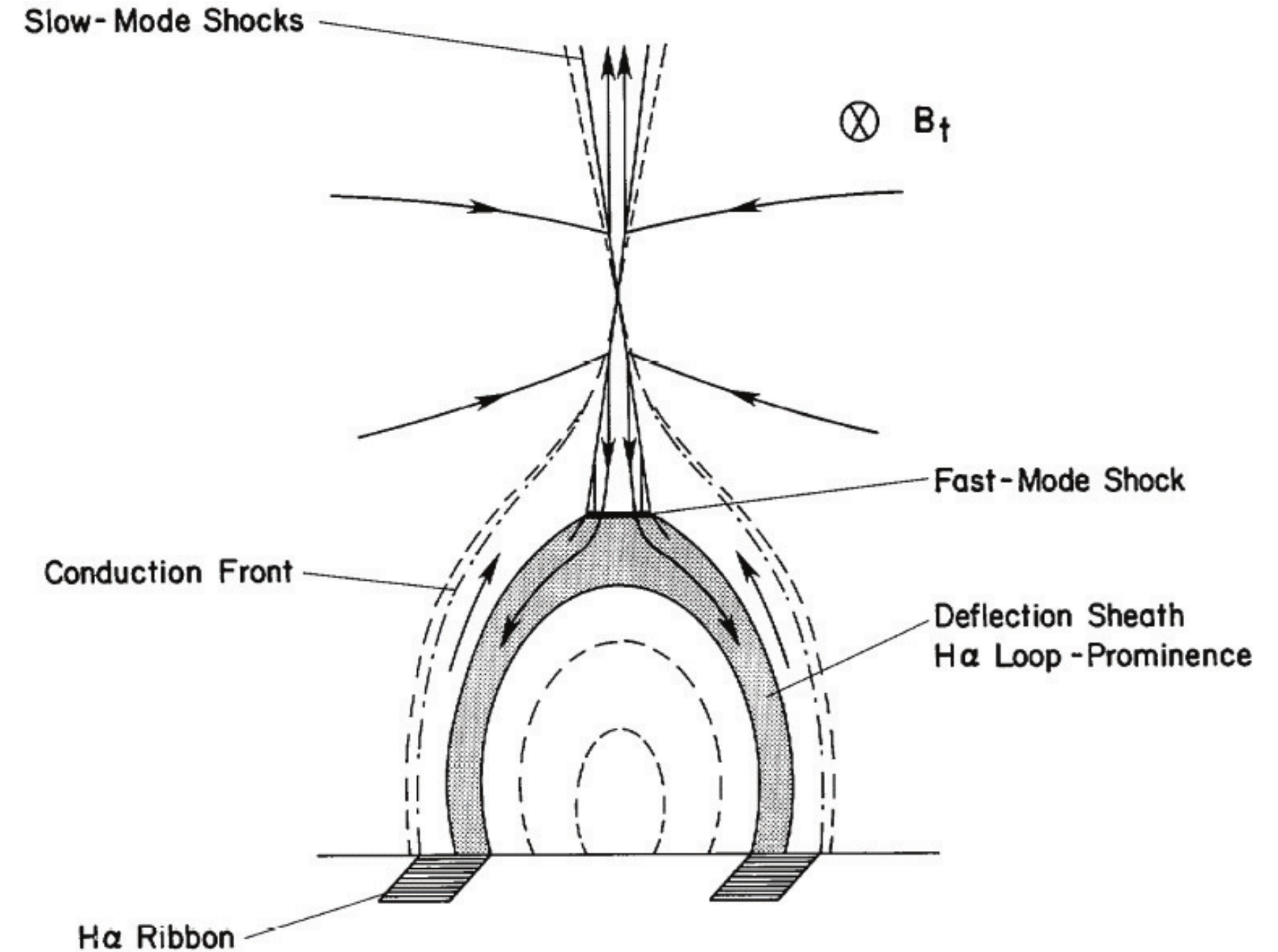}
\caption{A cartoon describing a version of the CSHKP model in which the upflow from the reconnection region becomes a CME and the downflow from the reconnection region leads to acceleration of energetic electrons at a shock front where downflow encounters the closed-field region. The explicit time-dependence is replaced by the flows at implicit boundaries of the model. (From \cite{Forbes1986}, reproduced by permission of the AAS.)}
\label{fig:CSHKP}
\end{figure}

Early versions of the ``standard'' model for a flare, also known as the CSHKP (initials of the authors of \cite{Carmichael1964,Sturrock1966,Hirayama1974,Knopp1976}) model, are two-dimensional (2D) with a bipolar magnetic field forming closed-loops below a vertical current sheet which separates regions of oppositely-directed nearly-vertical magnetic field (Figure~\ref{fig:CSHKP}). The magnetic energy release is described in terms of a stationary inflow of (frozen-in) magnetic field and plasma to the current sheet, where (2D) magnetic reconnection occurs, and a stationary outflow of reconnected magnetic field and plasma both up and down from the reconnection site. 

There is no explicit time-dependence in a CSHKP model, and hence there is no EMF. The time-dependence is implicit in the boundary conditions, involving inflows and outflows. The only electric field is a potential field, and the line-integral of this field along any closed path inside the boundary is zero. In 2D versions of the model the only allowed current is along the axis perpendicular to the 2D structures. Although the 2D assumption is relaxed in more recent 3D versions \cite{Janvier2014}, the flows remain stationary in such models, which therefore have no EMF. 

\vskip 5mm
\textbf{2.2  Circuit models}
\vskip 5mm

\begin{figure}
 \includegraphics[width=0.8\hsize]{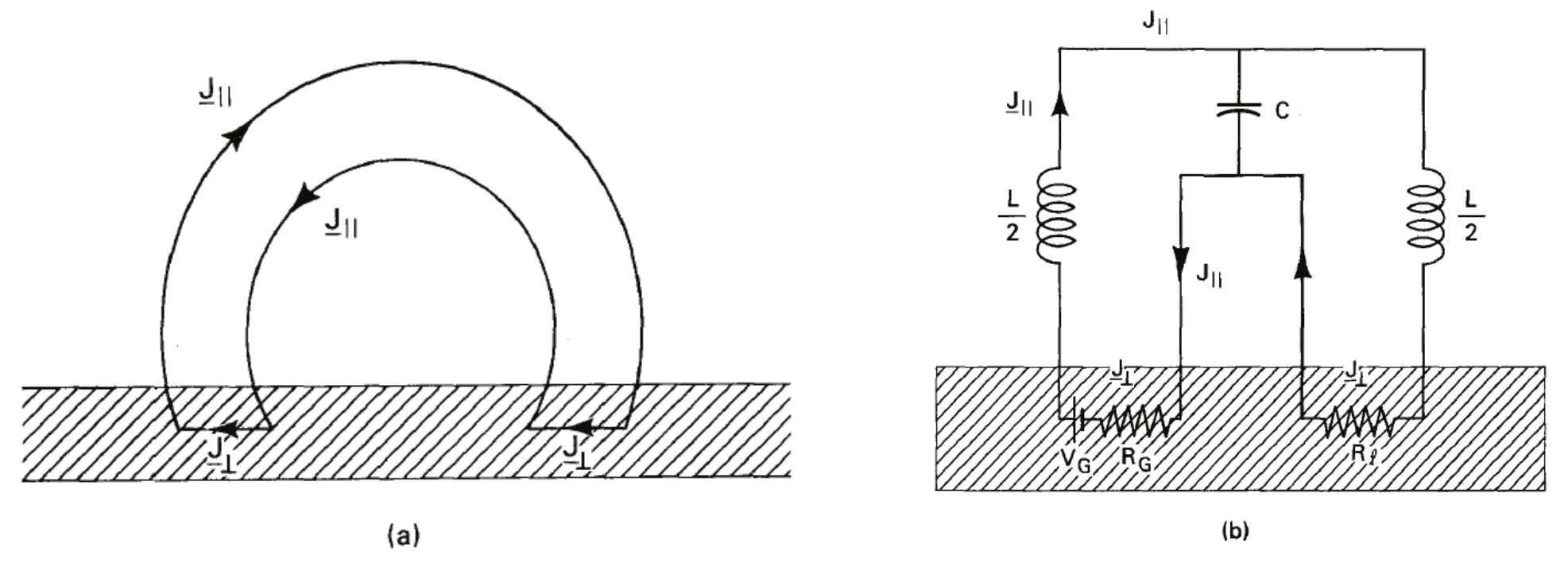}
\caption{A circuit model for a solar flare: left, assumed current path; right, circuit diagram showing inductance, $L$, capacitance, $C$, in the corona, and a voltage source $V_G$ and resistance in the (shaded) photosphere. (From \cite{S82}.)}
\label{fig:Spicer}
 \end{figure}

In a circuit model \cite{AC67,S82} the current is assumed to flow along a specified path, usually including a direct and return path along a coronal loop, with current closure across field lines at the two footpoints in the photosphere (Figure~\ref{fig:Spicer}).  Stored magnetic energy is identified as $\frac{1}{2}LI^2$, where $L$ is the inductance of the loop. The current is assumed to be driven by a voltage source located in one of the foopoints.  Thus, in this case the EMF is replaced by a photospheric dynamo. Also shown in Figure~\ref{fig:Spicer} is a capacitance, $C$: the kinetic energy in a mass motion is simulated by capacitive energy $Q^2/2C=C\Phi^2/2$, with the charge $Q$ building up due to the flare-associated current during the flare.

In the current-interruption version of a circuit model, the onset of a flare and the associated dissipation is attributed to the turning on of a coronal resistance, leading to ``interruption'' of the coronal current \cite{AC67}. Both direct and return current paths in the corona are needed to allow the postulated resistive region to short out the current between these two paths. Such a circuit/current-interruption model is not consistent with the electrodynamics of a magnetic explosion.

 \begin{figure}[htbp]
\centering
\includegraphics[width = 7.5cm]{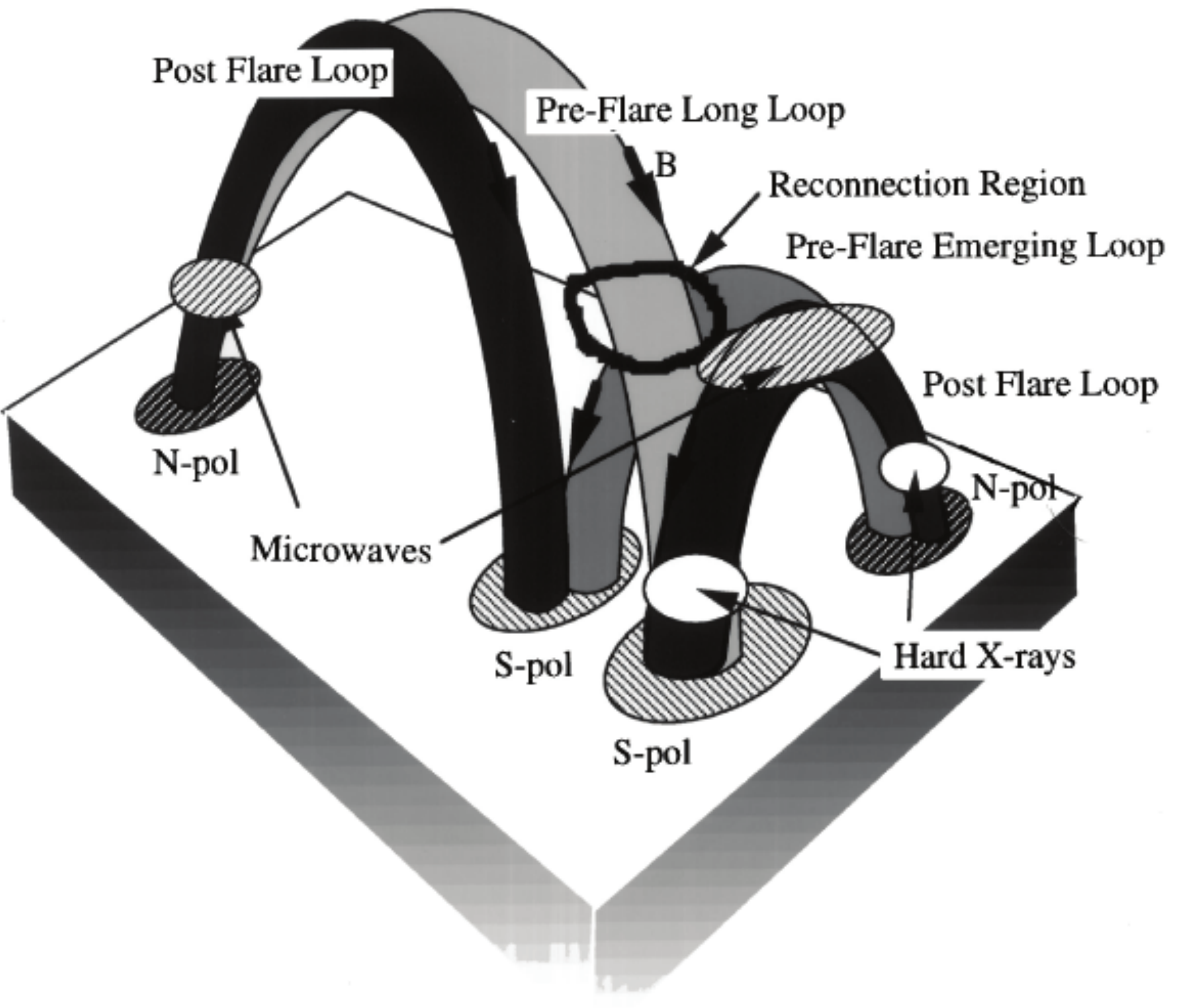}
\caption{A quadrupolar model for a flare showing two  (lightly shaded) pre-flare magnetic loops, two  (darkly shaded) post-flare loops, with the locations of the sources of microwaves and hard X-rays indicated. In the text, the North (N-pol) and South (S-pol) poles of the larger and smaller pre-flare  loops are labeled as footpoints $1\pm$ and $2\pm$, respectively, and the Reconnection Region is labelled $C$. (From \cite{Nishio1997}, reproduced by permission of the AAS.)}
\label{fig:nishio}
\end{figure}

\vskip 5mm
\textbf{2.3  Quadrupolar models}
\vskip 5mm

In a quadrupolar model \cite{Uchida1980,Nishio1997,Melrose1997,Zaitsev1998,Uchida1999} two bipolar magnetic loops come together and reconnect forming two new loops connecting the four footpoints (Figure~\ref{fig:nishio}). The reconnection between the magnetic fields in the two loops involves transfer of currents as well as magnetic flux between the loops. The release of magnetic energy is attributed to the change in the current paths, described by the change in the self- and mutual-inductances \cite{Melrose1997}, rather than a change in the currents. The energy release is estimated by comparing before and after states. Because the time-dependent changes during the flare are not considered, there is no EMF in these models. 

One way of including the time-dependence in such a model is by allowing the inductance to be time-dependent. The simplest example involves a single current-carrying flux tube that implodes \cite{Fletcher2008}, such that its length and hence its inductance decreases with time. In such a model the magnetic flux is $LI$, implying an EMF $-(dL/dt)I$ as $L$ decreases during the implosion. In a more general model involving multiple loops, the time-dependence is included in mutual inductances \cite{Khodachenko2009}.

\vskip 5mm
\textbf{3  Requirements for a time-dependent flare model}	
\vskip 5mm

A time-dependent flare model, driven by the EMF associated with the changing coronal magnetic field, requires several ingredients that are not present in most existing flare models. For the purpose of discussion order-of-magnitude estimates for $\Phi$ and $I$ are needed, and these are discussed in Sections~3.1 and~3.2, respectively.

\vskip 5mm
\textbf{3.1  EMF}	
\vskip 5mm

An early estimate of the EMF due to the time-varying magnetic field of a star \cite{Swann1933} gave $\Phi$ of order $10^{10}\,$V, and this was supported by later estimates for solar flares \cite{AC67,Colgate1978}. For example, an estimate of $\Phi$ from the rate of change of the magnetic flux during a flare is as follows. Equating the energy released, say $10^{24}\,$J in a moderate-sized flare, to the change in the magnetic flux times the current, which is estimated below to be of order $10^{11}\,$A, one infers that a change in flux of $10^{13}\,$Wb is needed. Such a change over $10^3\,$s implies an EMF of $10^{10}\,$V. There is an independent indirect argument that supports such a large $\Phi$ in an eruptive flare. The capacitor model for the kinetic energy of a CME involves equating the kinetic energy in the CME to the energy $Q^2/2C=C\Phi^2/2$ stored in the capacitor, also assumed to be  $10^{24}\,$J. With $C$ of order $10^4\,$F the required charge is $Q=10^{14}\,$C, and a current $I=10^{11}\,$A can build up this charge over the duration $10^3\,$s of a flare. The capacitive energy may also be written as $Q\Phi/2$, again suggesting $\Phi$ of order $10^{10}\,$V. Other variants of such estimates also lead to $\Phi$ of order $10^9$--$10^{10}\,$V. 

There is no direct signature for this enormous EMF, posing the question as to how it can seemingly be hidden in the corona during a flare. There is direct evidence from HXBs \cite{Brown1971,Brown1976} that a large fraction of the energy released goes into bulk energization of $\approx10$--$20\,$keV electrons. This suggests acceleration to an energy $\varepsilon=e\Delta\Phi$ with $\Delta\Phi$ of order $10^4\,$V. However, to explain a power of $10^{21}\,$W with an EMF $\Delta\Phi=10^4\,$V would require $I=10^{17}\,$A, which is impossible in the solar corona  \cite{Holman1985}. Assuming $I=10^{11}\,$A requires that $\Delta\Phi=\Phi/M$ appears across $M$ regions in series along the current path. A suggested geometry \cite{Holman1985} is $M$ pairs of up- and down-current paths with precipitating electrons being accelerated by $\Delta\Phi$ on each of the up-current paths. A global model for a flare needs to incorporate this requirement.

\vskip 5mm
\textbf{3.2  Flare-associated current}	
\vskip 5mm

The integrated form of Faraday's equation implies that the line-integral of the electric field along any path that encloses a changing magnetic flux is nonzero. However, most closed paths are irrelevant because a current cannot flow along them. The only relevant closed path is the one along which the flare-associated current actually flows, and this path needs to be identified. The EMF is the line-integral of the electric field along the actual path of the  flare-associated current. 

Energetically important coronal currents flow through the photosphere and cannot change significantly on the time scale of a flare \cite{Melrose1997}. The flare-associated current can be attributed to redirection or reconfiguration of currents that are flowing in the corona prior to the flare. It has long been known from vector magnetogram data \cite{MS68} that the currents flowing through the photosphere in a flaring region are of order $10^{11}$--$10^{12}\,$A. This suggests that appropriate fiducial values for a moderate-size flare are $\Phi=10^{10}\,$V and $I=10^{11}\,$A. With these values one can account for the power $I\Phi=10^{21}\,$W  and the energy $10^{24}\,$J released over $10^3\,$s.

This suggests that the flare-associated current is set up by Alfv\'en waves transferring a pre-flare current across field lines. Steady-state currents in the corona are nearly force-free implying that current lines are parallel to magnetic field lines. An Alfv\'en wave has a cross-field current, allowing transfer of a parallel current from one field line to another. A sequence of Alfv\'en waves, propagating between two end points can set up a closed-loop current over a time scale of many Alfv\'en propagation times between the end points \cite{Melrose1992}.  However, this is possible only if the (quasi-steady-state) current can close across field lines at the two ends. The current flows in the opposite direction to the initial current along one field line, thereby partially canceling the initial current, and in the direction of the current along another field line which becomes the final current path. The flare-associated current may involve several such closed-loop currents in the corona. It also includes additional current loops set up between a reconnection region in the corona and a current-closure region in the chromosphere. With the flare-associated current due to reconfiguration of pre-existing currents, it follows that the (maximum) flare-associated current should be comparable in magnitude with the current flowing in the coronal flux tube prior to the flare, assumed here to be $I=10^{11}\,$A. 

Currents can flow freely along field lines in the corona, but can flow across field lines only under special conditions. In a specific model for the flare-associated current the special conditions that allow cross-field closure at the two ends need to be identified. There are three possibilities for such cross-field closure.
\begin{itemize}
\item A cross-field current implies a ${\bf J}\times{\bf B}$ force, which must be balanced by another force or associated with the acceleration of mass. Only the latter seems plausible for the Lorentz force due to the flare-associated current in the low-beta plasma in the solar corona. 
\item A cross-field current can flow at a footpoint of a coronal loop, due to the cross-field electrical conductivity (Pedersen conductivity) of the (partially ionized) chromospheric plasma. Due to the large inertia of the chromosphere, such current closure may lead to only a small acceleration of plasma. Although the finite conductivity implies some dissipation, this is unimportant in the overall energy budget for a flare.
\item The third possibility may be called \textit{reconnection-associated current closure}. When current-carrying flux tubes intersect and reconnect, both magnetic flux and (field-aligned) current are partly transferred to two new flux tubes. The field-aligned current paths are changing as a function of time during such reconnection, complicating the classification into field-aligned and cross-field currents.
\end{itemize}
In a single-loop model there are two regions of cross-field current closure. One region of cross-field closure is in the corona where the Lorentz force produces mass motion driven by the flare. An argument invoked for both a laboratory plasma \cite{Simon1955} and for the terrestrial magnetosphere during a substorm \cite{McPherron1973} suggests that the other cross-field closure is at a footpoint; in a flare this corresponds to closure due to the cross-field (Pedersen) conductivity in the chromosphere. 

For example, in a quadrupolar model  in Figure~\ref{fig:nishio} reconnection-associated current closure in the corona is assumed to occur at $C$. The net effect of the flare-associated current must be to change the initial current configuration to the final current configuration. This quadrupolar model requires (partial) cancelation of the currents from $1+$ to $1-$ and from $2+$ to $2-$ and creation of currents from $1+$ to $C$ to $2-$ and from $2+$ to $C$ to $1-$, where $C$, $1\pm$ and $2\pm$ are defined in the caption to Figure~\ref{fig:nishio}. In this case two flare-associated current paths are needed, one from $1-$ to $C$ to $2+$ to $1-$ and the other from $2-$ to $C$ to $1+$ to $2-$. The newly formed currents are transferred to their post-flare locations, as illustrated in Figure~\ref{fig:nishio}, as the (reconnected) magnetic field lines along which the currents flow move to their post-flare locations.

The flare-associated current must pass through acceleration/dissipation regions, where the magnetic energy released in the corona is transferred to energetic electrons. The source of the energetic electrons and the locations of the acceleration/dissipation regions are constrained by the ``number problem''.

\vskip 5mm
\textbf{3.3  Number problem}	
\vskip 5mm

A long-standing difficulty in interpreting hard X-ray bursts (HXBs) in solar flares is that the number of electrons precipitating into the denser regions of the solar atmosphere, where HXBs are generated, greatly exceeds the number of electrons in the flaring flux tube prior to the flare \cite{Brown1971,Brown1976,MacKinnon1989}. Electrons must be resupplied to the corona at the same rate as energetic electrons precipitate from the corona. This requirement can be satisfied by invoking a return current \cite{Hetal76}, involving electrons flowing up from the chromosphere into the corona. Early models for the return current \cite{Knight1977, BM77,BB84,Spicer1984,vandenOord1990} invoked co-located direct and return currents. An alternative is that the electrons drawn up from the chromosphere at one footpoint of a flaring flux tube are accelerated along their path before precipitating at the conjugate footpoint \cite{Emslie1995}.

An up-current of $I=10^{11}\,$A at the photosphere corresponds to electrons flowing down at $I/e\approx10^{30}\rm\,s^{-1}$. The thick-target model for HXBs \cite{Brown1971} requires energetic electrons precipitating at a rate ${\dot N}$ of order $10^{36}\rm\,s^{-1}$ or greater. The two rates, $I/e$ and ${\dot N}$ differ by a factor of the same order as the mismatch $M=\Phi/\Delta\Phi$ between the energy $e\Phi$ and the typical energy $\varepsilon=e\Delta\Phi$ of order $10^4\,$eV of the energetic electrons that generate HXBs and type~III bursts. Both factors are of order $10^6$. This approximate equality implies that the power $I\Phi$, estimated from the electrodynamics, and the power in precipitating electrons, ${\dot N}\varepsilon$, are of the same order of magnitude. A model is needed to provide an interpretation as to why $M$ is of order $10^6$.

\vskip 5mm
\textbf{3.4  Hiding the ``elephant in the room''}	
\vskip 5mm

An explanation of the mismatch $M$ of order $10^6$ involves several different ingredients. One ingredient is the assumption that the direct and return currents are set up in pairs through propagation of Alfv\'enic fronts between the regions of cross-field current flow in the corona and in the (partially ionized) chromosphere \citep{MW13}. The Alfv\'en waves transport energy (and potential) downward from the corona along field lines. Dissipation is essential in providing a sink for this energy. If there is no dissipation there is no release of magnetic energy, no EMF and no energy transport. At the start of the flare, the dissipation is required to build up allowing magnetic energy release to build up, and $\Phi$ to increase. A second ingredient is that there is a maximum $\Phi_{\rm max}$ that can be supported by Alfv\'en waves \citep{MW14}. Once  $\Phi$ reaches $\Phi_{\rm max}$, any further increase in the rate of energy release, energy transport and energy dissipation requires a further pair of direct and return currents to develop in series with the first. This results in many pairs of direct and return currents that develop in series, with $\Delta\Phi\approx\Phi_{\rm max}$ across field lines between direct and return currents in each pair. Such multiple pairs of current paths has been described as a ``picket-fence'' model \citep{MW13}. 

\vskip 5mm
\textbf{3.5  Reconnection region}
\vskip 5mm 

Magnetic reconnection is an essential ingredient in any model for a magnetic explosion: it is required to allow the magnetic configuration to change. During a flare, current-carrying magnetic field structures in the corona must change from an initial configuration with a higher stored magnetic energy to a final configuration with a lower stored magnetic energy. Although there have been calculations of reconnection involving twisted flux ropes \cite{Linton2001}, how such models relate to the magnetic and current configuration on a global scale, for example as illustrated in Figure~\ref{fig:nishio}, is unclear. Magnetic reconnection in the solar corona can occur only in localized regions, usually assumed to be associated with magnetic nulls and other special structures \cite{Longcope2005}. Compared with a solar flare on a global scale, the spatial and temporal scales of a reconnection event are microscopic, and a statistically large number of localized reconnection regions is needed to have a macroscopic effect. One model involving many magnetic nulls is referred to as turbulent reconnection \cite{Lazarian1999}. 

As in a CSHKP model, magnetic energy is assumed to flow into the reconnection region in the form of frozen-in magnetic flux. In a time-dependent model, the rate of change of magnetic energy density, $-\partial(B^2/2\mu_0)/\partial t$, is locally balanced by the divergence of the energy flux $(B^2/\mu_0){\bf u}$, where ${\bf u}$ is the fluid velocity. This inflow builds up over a large volume around the reconnection region. The magnetic energy is partly converted into other forms in the reconnection region, and the energy inflow is balanced by an outflow that includes kinetic energy, energetic particles and heat. However, while this general description applies to most models that involve reconnection, the role played by the reconnection region is different in different models. 

In a CSHKP model, the reconnection region is also identified as the region where the conversion of energy into mass motion in a CME and into energetic particles occurs. In early CSHKP models, a CME was identified as an upward fast outflow from the reconnection region. The acceleration of particles was assumed to occur either during the reconnection itself, or at a shock wave where the downward fast outflow encounters the underlying closed-field region (Figure~\ref{fig:CSHKP}). Although these features of a CSHKP model are qualitatively appealing, such a model is no longer favored for the generation of a CME, and it encounters seemingly insurmountable quantitative difficulties (the ``number problem'') when compared with data on HXBs. 

A different interpretation of the energetics of the reconnection region is adopted here. It is assumed that the energy conversion in the reconnection region is the first stage in a multi-stage energy conversion process. The second stage includes an energy outflow from the reconnection region, as a Poynting flux in Alfv\'enic form. There is observational evidence for such a model for geomagnetic substorms \cite{Ergun2002,Chaston2002}, and a similar model has been suggested for solar flares \cite{Haerendel1994,Haerendel2012,Fletcher2008}. Although no detailed model for the conversion of the inflowing magnetic energy into an outflowing Alfv\'en flux is available, some features of such a model are evident. The flare-associated current flowing along field lines in the corona must be redirected at the reconnection region, so that the current flows along field lines to and from the chromosphere, where it flows across field lines due to the Pedersen conductivity. There is a cross-field electric field, ${\bf E}_\perp$, between the direct and return current paths, and this may be attributed to the EMF becoming localized across the reconnection region. (Specifically, $E_\perp=\Delta\Phi/L_\perp$, where $L_\perp$ is the distance between the up- and down-current paths.) The magnetic field due to the up- and down-currents combined with ${\bf E}_\perp$ gives the Poynting flux in Alfv\'enic form. The cross-field potential is also transported towards the chromosphere. As in the acceleration of auroral electrons in a geomagnetic substorm, it is assumed that an acceleration/dissipation region develops on the up-current path, with electrons drawn up from the chromosphere in the downward or return current path. 

Such a model involves magnetic energy release in at least three stages. The first stage is the conversion of magnetic energy into a Poynting flux that gives the energy inflow into the reconnection region. The second stage is the energy outflow from this region into a downward Alfv\'enic Poynting flux towards the chromosphere. The third stage is the conversion of this Alfv\'enic energy flux into energetic electrons in the acceleration/dissipation region. A further stage is needed to explain the energy transfer to a CME in an eruptive flare.

\vskip 5mm
\textbf{4  Acceleration/dissipation region}
\vskip 5mm

A long-standing problem in the physics of solar flares is the ``bulk energization'' of the energetic (10--$20\,$keV) electrons that produce HXBs and type III solar radio bursts in the impulsive phase \cite{Ramaty1980,Benz1987,Melrose1990}. The location of the acceleration region, and the acceleration mechanism are different in different models. 


A CSHKP model favors acceleration along the neutral plane separating oppositely directed magnetic field lines, cf.\ Figure~\ref{fig:CSHKP}. One suggestion is that the electron acceleration is associated with contracting magnetic islands during reconnection \cite{Drake2006}. Such acceleration involves a parallel electric field, $E_\parallel$, in the collapsing magnetic island. However, the number problem provides a strong argument against this being the bulk-energization mechanism: it seems impossible for such a model to account for the large number of electrons accelerated. However, this is an argument against the specific model, rather than against acceleration by $E_\parallel$ in general. The quantitative argument favors acceleration in or near the chromosphere, where there is a copious supply of electrons. Other arguments in favor of acceleration relatively low in the solar atmosphere include a suggested analogy with the acceleration of auroral electrons in a geomagnetic substorm \cite{Haerendel1994,Haerendel2012}, an Alfv\'en-wave model for energy transport \cite{Fletcher2008} and observational evidence for the number density of the precipitating electrons being of order the ambient number density \cite{Krucker2011}.

The fact that the details of the acceleration mechanism remain uncertain is not surprising: acceleration by $E_\parallel$ is poorly understood in any space or astrophysical context. In particular, the acceleration of auroral electrons is attributed to the potential $\Delta\Phi$ changing from across field lines to along field lines, producing $E_\parallel$ \cite{Ergun2002}. However, despite the acceleration region being probed by spacecraft, the detailed plasma physics remains inadequately understood. Various specific models for $E_\parallel\ne0$ have been suggested including double layers, anomalous conductivity and phase-space holes \cite{Main2006}. Although it is desirable that the micro\-physics be understood, there is an argument that suggests that a detailed understanding may not be essential. Each of the structures suggested for $E_\parallel\ne0$ is on a very small spatial and temporal scale, and a statistically large number of such structures is needed to have a macroscopic effect. Such a statistical model, perhaps based on network theory or some related model, may be insensitive to the microphysics and allow the acceleration/dissipation to be described in terms of a macroscopic quantity, such as an equivalent resistance.
	
\vskip 5mm
\textbf{5 Discussion and conclusions}	
\vskip 5mm

As suggested earlier \cite{Melrose2012}, it is important to introduce widely different scales in order to model magnetic energy release in a solar flare. It is on the largest (``global'') scale that our thinking about flares needs to be revised to include time-dependent electrodynamics explicitly: specifically, the time-changing magnetic field, the EMF and the flare-associated current, with the release of magnetic energy due to the work done by the EMF against the flare-associated current. There are several questions that need to be answered  but which are obscured by the neglect of the EMF and the flare-associated current in most flare models, as discussed in Section~2. One question is how the very large EMF, $\Phi$ of order $10^{10}\,$V, can apparently be hidden in the corona: the suggested answer given here is essentially as originally proposed by Holman in 1985 \cite{Holman1985}. Answers to other questions related to the flare-associated current suggest that it is due to redirection of pre-flare coronal currents, and that it involves more than one current loops in the corona, and multiple ($M$) pairs of up- and down-current paths between a reconnection region and the current closure region in the chromosphere. Answers to further questions relating to energy transport and energy dissipation (through bulk energization) are suggested. Although the detailed answers to such questions are important, the main point made here is that a new way of thinking about flares is needed in order to recognize that such questions are relevant and to address them. This  new way of thinking must be based on time-dependent electrodynamics. 

\begin{acknowledgments}
I thank Alpha Mastrano and Mike Wheatland for helpful comments on the manuscript. I also acknowledge helpful discussions with members of the team on ``Magnetic Waves in Solar Flares'' at the International Space Science Institute, Bern, Switzerland.
\end{acknowledgments}

\end{CJK}

\end{document}